\documentclass[conference]{IEEEtran}

\usepackage{algorithmic}
\usepackage{graphicx}
\usepackage{textcomp}
\usepackage[dvipsnames]{xcolor}
\usepackage{xspace}
\usepackage{fontawesome}
\usepackage{multirow}
\usepackage{colortbl}
\usepackage{rotating}
\usepackage{tikz}
\usepackage{soul}
\usepackage{makecell}
\usepackage{url}
\usepackage{enumitem}
\usepackage{fancybox}
\usepackage{booktabs}
\usepackage{fontawesome}
\usepackage{setspace}               
\usepackage{pifont}
\usepackage[many]{tcolorbox}    	
\usepackage[caption=false]{subfig}
\usepackage[export]{adjustbox}
\usepackage{amsmath,amsfonts}
\usepackage{MnSymbol}
\usepackage{color,colortbl,booktabs}
\usepackage[noadjust]{cite}
\usepackage{balance}

\definecolor{main}{HTML}{cccccc}    
\definecolor{sub}{HTML}{000000}     

\newtcolorbox{boxM}{
	fontupper = \color{black},
	rounded corners,
	arc = 6pt,
	colback = main!80, 
	colframe = main, 
	boxrule = 0pt, 
	bottomrule = 4.5pt,
	enhanced,
	fuzzy shadow = {0pt}{-3pt}{-0.5pt}{0.5pt}{black!35}
}

\newcommand{\ie}{\emph{i.e.,}\xspace}
\newcommand{\eg}{\emph{e.g.,}\xspace}

\newcommand{\etal}{\emph{et~al.}\xspace}
\newcommand{\secref}[1]{Section~\ref{#1}\xspace}

\newcommand{\figref}[1]{Fig.~\ref{#1}\xspace}

\newcommand{\tabref}[1]{Table~\ref{#1}\xspace}

\newcommand*\circled[1]{\tikz[baseline=(char.base)]{
		\node[shape=circle,fill,inner sep=0.8pt] (char) {\textcolor{white}{#1}};}}
\newboolean{showcomments}
\setboolean{showcomments}{true}

\ifthenelse{\boolean{showcomments}}
{\newcommand{\nb}[2]{
		\fbox{\bfseries\sffamily\scriptsize#1}
		{\sf\small$\blacktriangleright$\textit{#2}$\blacktriangleleft$}
	}
	
}
{\newcommand{\nb}[2]{}
	
}

\AtBeginDocument{%
	\providecommand\BibTeX{{%
			Bib\TeX}}}
\def\BibTeX{{\rm B\kern-.05em{\sc i\kern-.025em b}\kern-.08em
    T\kern-.1667em\lower.7ex\hbox{E}\kern-.125emX}}
\begin{document}

\title{Resource-Efficient \& Effective Code Summarization}

\author{
	\IEEEauthorblockN{Saima Afrin\IEEEauthorrefmark{1}, Joseph Call\IEEEauthorrefmark{2}, Khai-Nguyen Nguyen\IEEEauthorrefmark{3}, Oscar Chaparro\IEEEauthorrefmark{4}, Antonio Mastropaolo\IEEEauthorrefmark{5}}
	\IEEEauthorblockA{
		\textit{William \& Mary}, \textit{Department of Computer Science} \\
		\textit{Williamsburg, Virginia, USA} \\
		\IEEEauthorrefmark{1}safrin@wm.edu,
		\IEEEauthorrefmark{2}jbcall@wm.edu,
		\IEEEauthorrefmark{3}knguyen07@wm.edu,
		\IEEEauthorrefmark{4}oscarch@wm.edu,
		\IEEEauthorrefmark{5}amastropaolo@wm.edu}
}

%
%
%
%
%
%
%
%

\maketitle

\begin{abstract}
Code Language Models (CLMs) have  demonstrated high effectiveness in automating software engineering tasks such as bug fixing, code generation, and code documentation. This progress has been driven by the scaling of large models, ranging from millions to trillions of parameters (\eg GPT-4).
However, as models grow in scale, sustainability concerns emerge, as they are extremely resource-intensive, highlighting the need for efficient, environmentally conscious solutions. \textcolor{ForestGreen}{GreenAI} techniques, such as QLoRA (Quantized Low-Rank Adaptation), offer a promising path for dealing with large models' sustainability as they enable resource-efficient model fine-tuning.
Previous research has shown the effectiveness of QLoRA in code-related tasks, particularly those involving natural language inputs and code as the target output (NL-to-Code), such as code generation. However, no studies have explored its application to tasks that are fundamentally similar to NL-to-Code (natural language to code) but operate in the opposite direction, such as code summarization. This leaves a gap in understanding how well QLoRA can generalize to Code-to-NL tasks, which are equally important for supporting developers in understanding and maintaining code.
To address this gap, we investigate the extent to which QLoRA's capabilities in NL-to-Code tasks can be leveraged and transferred to code summarization, one representative Code-to-NL task.
Our study evaluates two state-of-the-art CLMs (CodeLlama and DeepSeek-Coder) across two programming languages: Python and Java. Each model was tasked with generating a meaningful description for Python and Java code methods. The findings of our research confirm previous patterns that emerged when applying QLoRA to source code generation. Notably, we observe that QLoRA not only allows efficient fine-tuning of CLMs for code summarization but also achieves the best results with minimal parameter adjustment compared to full model fine-tuning, which requires expensive recalibration of all model parameters in the traditional fine-tuning process.
\looseness=-1

\end{abstract}


\section{Introduction}
\label{sec:intro}

In recent years, deep learning (DL) generative models, particularly Large Language Models (LLMs) and  Code Language Models (CLMs), have transformed key software engineering (SE) activities, including bug fixing, code generation, and code documentation~\cite{charalambous2023new,mastropaolo2021studying,mastropaolo2024evaluating,tian2023chatgpt,zhang2020retrieval}.
These advances have significantly enhanced automation, driving productivity in SE workflows.

To fully realize their potential, CLMs often require fine-tuning to achieve high accuracy on specific tasks.
Prior research~\cite{weyssow2023exploring,liu2022few} has shown that fine-tuned CLMs outperform pre-trained CLMs that rely on in-context learning (ICL), particularly for downstream tasks such as code summarization~\cite{ahmed2024automatic,ahmed2022few,wang2022no}.
Fine-tuning allows for deeper calibration of model parameters, resulting in higher adaptability \& robustness for specific tasks.
\looseness=-1

However, fine-tuning large-scale language models—often comprising billions of parameters—demands significant computational resources and time \cite{hou2023large}. For instance, training the CodeLlama \cite{codellama} family of models reportedly required over 1.4 million GPU hours \cite{codellama2}, highlighting the substantial effort needed to achieve state-of-the-art performance.

In response to this challenge, researchers have explored sustainable methods to reduce the environmental and computational costs associated with training large-scale models while maintaining high performance \cite{wei2023towards, weyssow2023exploring, ayupov2022parameter, lu2023llama}. Techniques such as model compression and parameter-efficient fine-tuning (PEFT) \cite{ayupov2022parameter, lu2023llama, weyssow2023exploring, su2024distilled, shi2023towards} have emerged as promising solutions, enabling efficient training with significantly lower resource demands.

One recent advancement at the intersection of model compression and PEFT is QLoRA (Quantized Low-Rank Adaptation)~\cite{dettmers2024qlora}, a technique that combines model size reduction with efficient fine-tuning strategies. QLoRA has been shown to enable cost-effective fine-tuning of CLMs for tasks such as program repair and code generation/completion~\cite{weyssow2023exploring, yang2024multi}, which fall into the categories of Code-to-Code and NL-to-Code (natural language to code) tasks. These results suggest that QLoRA significantly reduces computational overhead while achieving high effectiveness compared to methods requiring full parameter calibration. Despite these promising results, the applicability of QLoRA to Code-to-NL tasks, such as code summarization, remains unknown. This paper addresses this gap by evaluating QLoRA's effectiveness for code summarization.
\looseness=-1

Code summarization, like other bi-modal code-related tasks (\eg code review and code generation), requires reasoning across code and natural language, with the aim to translate complex code logic into accurate, clear, and concise natural language explanations. Given that QLoRA has proven effective for code generation~\cite{weyssow2023exploring}, we \textit{hypothesize} that it is equally effective for code summarization. This hypothesis is grounded in the conceptual parallel between teaching a model to generate code and teaching it to summarize code, as both tasks involve an inverse relationship where input and output roles are reversed, with both tasks learning nuanced relationships between natural and programming languages.

To validate this hypothesis, we conducted a systematic evaluation of QLoRA using two state-of-the-art CLMs, CodeLlama~\cite{codellama} and DeepSeek-Coder \cite{deepseek}, designed to summarize code methods written in Python and Java from the CodexGLUE's code summarization dataset\footnote{\url{https://tinyurl.com/axbp8hua}}. We trained these models with QLoRA under varying parameter sizes and compared their performance to full model fine-tuning, analyzing memory usage and predictive accuracy. Additionally, we qualitatively analyzed two statistically significant samples of code methods—one comprising Python methods and the other Java methods—to evaluate how closely the generated summaries align with the ground truth and how effectively they convey equivalent information. This analysis establishes a virtual upper bound on the potential effectiveness of QLoRA for Code-to-NL tasks, particularly code summarization.


Our results show that QLoRA achieves superior predictive performance compared to full fine-tuning while consistently reducing the memory footprint of CLMs. These findings provide compelling evidence of QLoRA's ability to optimize CLMs for resource-intensive, bi-modal code-related tasks, thereby showing its utility across the full spectrum of code-related tasks: Code-to-Code, NL-to-Code, and  Code-To-NL.

To the best of our knowledge, this work represents the first large-scale evaluation of QLoRA for code summarization, and it makes the following key contributions:

\begin{itemize}
	\item A comprehensive analysis of QLoRA’s capabilities for code summarization, using two state-of-the-art CLMs across two programming languages, contributing to a broader understanding of resource-efficient training across the full spectrum of code-related tasks.

	\item Key insights into the trade-offs between memory usage and model performance compared to full model fine-tuning, showcasing QLoRA's ability to achieve remarkable results with substantially reduced resource requirements in the context of code summarization.
	
	\item A replication package~\cite{replication}, including data, models, scripts, and documentation, to facilitate reproducibility and further research in this field.
\end{itemize}

\section{Background and Related Work}
\label{sec:related}

This section provides the reader with an overview of recent advancements in efficiency-based methods that aim to improve the sustainability of large language models, particularly code language models (CLMs) for code summarization.

\subsection{Code Language Models in Code Summarization} 
\label{sec:CLM}

Given the significant potential of SE-related automation through large code models grounded on LLMs, researchers increasingly leveraged these models to support various tasks, including those requiring higher levels of abstraction. One such task is code summarization, which involves working with bi-modal data to translate and summarize code into natural language. In this task, LLMs have proven highly effective~\cite{sun2024extractive,shi2022evaluation,fang2024esale,ahmad2020transformer,mastropaolo2022using}. CLMs like Codex \cite{ahmed2023improving,arakelyan2023exploring}, CodeBERT \cite{chen2022transferability, gu2022assemble}, and T5 \cite{mastropaolo2022using} excel in understanding code functionality and logic, generating clear and concise summaries. For example, Mastropaolo \etal~\cite{mastropaolo2022using} pre-trained a T5-based model on a blend of code and technical natural language before fine-tuning it on various code-related tasks, including code summarization. Their results highlighted the advantages of leveraging transfer learning for bi-modal code-related tasks, particularly code summarization. Haldar \etal~\cite{haldar2024analyzing} investigated the use of CodeT5 \cite{wang2021codet5}, PaLM2 \cite{anil2023palm}, and Llama2 \cite{touvron2023llama} to generate meaningful code summaries. While CodeT5 was subject to fine-tuning, PaLM2 and Llama2 required no parameter adjustment. The authors' findings reveal that LLMs frequently leverage function names and shared tokens between the code and its summary to optimize predictive performance.

Ahmed \etal \cite{ahmed2022few} found that few-shot prompting, which involves providing the model with few examples for generation tasks, significantly improves Codex's performance in code summarization, outperforming smaller pre-trained models like CodeT5. In another study,  Sun \etal \cite{sun2023automatic} explored CodeLlama~\cite{codellama2} and GPT-4 \cite{achiam2023gpt} for code summarization and evaluated five prompting techniques (\ie zero-shot, few-shot, chain-of-thought, critique, and expert). They identified the most effective prompt for guiding  GPT-4  to generate in-distribution code summaries.

\subsection{Parameter-Efficient Fine-Tuning and Quantization Methods}
\label{sec:peft-qt}
\textit{Parameter-Efficient Fine-Tuning} (PEFT) optimizes fine-tuning by updating only a subset of a model's parameters, rather than the entire model. Common techniques include: (i) Adapters, where additional model layers are introduced to handle a limited set of parameters \cite{houlsby2019parameter}; (ii) Prompt Tuning, which trains the model to learn from prompts containing task descriptions or canonical examples \cite{lester2021power, li2021prefix}; and (iii)
LoRA (Low-Rank Adaptation), which  decomposes weight gradients into low-rank matrices during fine-tuning~\cite{hu2021lora}. 
\looseness=-1
 
 PEFT has shown strong performance in tasks such as code generation and summarization, often outperforming fully fine-tuned models. For instance, Wang \etal~\cite{wang2023one} applied Adapter tuning for code search and summarization, while Ayupov~\etal~\cite{ayupov2022parameter} showcased the effectiveness of Adapters and LoRA  in tasks like code summarization and code clone detection. Similarly, Liu \etal~\cite{liu2023empirical} compared PEFT methods—such as Adapter, LoRA, prefix tuning, and Multi-Head Modification~(MHM)—for tasks like defect detection, clone detection, code translation, and code summarization. Recent studies~\cite{sun2023prompt, shi2023towards} have further explored PEFT techniques in the context of code summarization, highlighting their importance in this domain.
 \looseness=-1

\textit{Quantization} is a technique for model compression. It aims to reduce the size of a model by preserving only the most essential information encoded in the model's parameters. Specifically, it achieves compression by representing weights or activations in lower-precision formats, such as 8-bit integers, rather than higher-precision formats like 16-bit or 32-bit floats~\cite{gholami2022survey, zhu2023survey}. This approach reduces latency while minimizing any potential loss in accuracy.

\setlength{\parskip}{1pt}



In the software engineering domain, the pioneering study by Wei \etal~\cite{wei2023towards} represents the first large-scale investigation into the application of quantization techniques for code-related tasks, including code generation and summarization. The authors examined the effects of 8-bit quantization on various code models, such as PLBART \cite{ahmad2021unified}, CodeT5 \cite{wang2021codet5}, InCoder \cite{fried2022incoder}, and CodeGen \cite{nijkamp2022codegen}. Their findings revealed that applying 8-bit quantization to CodeGen and InCoder resulted in improved energy efficiency during code generation, while PLBART and CodeT5 showed similar benefits for code summarization. Notably, these gains in efficiency were achieved with only a minimal reduction in model accuracy.

\subsection{Quantized Low-Rank Adaptation (QLoRA) of CLMs}
\label{sec:qlora}

Dettmers \etal~\cite{dettmers2024qlora} recently proposed QLoRA, an approach that combines the LoRA PEFT technique with quantization of LLMs. QLoRA introduces various key innovations, including (i) the 4-bit NormalFloat (NF4) data type, (ii) Double Quantization (DQ), and (iii) a Paged Optimizer. It has been shown to be an efficient fine-tuning method that reduces memory usage while preserving the high performance of LLMs~\cite{dettmers2024qlora}. QLoRA quantizes the pre-trained model’s weights to 4-bit precision using NF4, a data type optimized for the normal distribution of neural network weights. Additionally, through double quantization, both the model weights and the quantization constants are quantized, further reducing the memory footprint. To manage memory spikes during gradient checkpointing and prevent out-of-memory errors, QLoRA employs Paged Optimizers. A detailed explanation of QLoRA and the fine-tuning process to achieve its goals is provided in \secref{sub:design_qlora}.

Limited research has investigated the efficiency of QLoRA for code language models. Yang \etal~\cite{yang2024multi} applied QLoRA on models such as CodeLlama \cite{codellama2}, StarChat-alpha \cite{Tunstall2023starchat-alpha}, and Mistral-Instruct-7B \cite{mistral} to specialize large code models for automatic program repair (APR). Their findings demonstrate that QLoRA effectively supports LLMs in repairing defects in software systems. Weyssow \etal~\cite{weyssow2023exploring} compared PEFT techniques to In-Context Learning (ICL) for code generation, concluding that PEFT methods achieved superior results. In addition, the authors also investigated the applicability of QLoRA to CodeLlama 7B, 13B, and 34B Python models, using 8-bit and 4-bit quantization. 

While these findings provide valuable insights, a comprehensive evaluation of whether QLoRA can effectively support the entire spectrum of code-related tasks--namely NL-to-Code, Code-to-Code, and Code-to-NL--remains absent. To address this gap, this paper takes a significant first step toward exploring QLoRA’s potential across these task categories. Specifically, we focus on Code-to-NL tasks, using code summarization as a representative case study, to evaluate how well QLoRA adapts in scenarios where the model processes code as input and generates natural language as output. 
This work seeks to deepen the understanding of resource-efficient training methods in software engineering tasks while providing a foundation for future research across diverse bi-modal tasks (\eg code review automation).

\section{Study Methodology} 
\label{sec:design}


The main goal of this study is to investigate the application QLoRA fine-tuning to code language models (CLMs) for code summarization. QLoRA combines PEFT and quantization techniques, resulting in substantial improvements in memory efficiency during LLM training compared to LoRA~\cite{dettmers2024qlora}. The study addresses the following research question (RQ): 

\begin{itemize}[label=,leftmargin=0.2cm]
	\item \textbf{RQ:} \emph{How effective and memory-efficient are CLMs for code summarization when fine-tuned with QLoRA, compared to full fine-tuning?}

\end{itemize}

Through this RQ, we aim to validate our hypothesis that QLoRA is equally effective for code summarization as it is for code generation~\cite{weyssow2023exploring}.

To answer the RQ, we examine two state-of-the-art code models: CodeLlama~\cite{codellama} and DeepSeekCoder \cite{deepseek}. Each model is trained and evaluated on the CodexGLUE code summarization benchmark \cite{codexglue}, particularly the dataset comprising Python and Java code methods and their respective summaries~\cite{CodeXGLUEbench}. 
\looseness=-1

Additionally, we investigate the impact of scaling up the parameters of CLMs during QLoRA-based training, measuring changes in GPU memory usage and overall predictive performance for code summarization. This analysis aims to determine whether larger models retain their performance advantage, as demonstrated in previous studies \cite{sun2024extractive,shi2022evaluation,fang2024esale,gros2020code}, where increasing the number of model parameters has consistently improved task-specific performance.

We also investigate the generalizability of QLoRA for LLMs that, while widely utilized for automating SE-related tasks, were not primarily designed for such tasks. The details of this analysis are provided in \secref{sub:phi3}.


\subsection{Code Language Models (CLMs)} 
\label{sub:design_llm}

For our study, we selected two families of state-of-the-art CLMs: CodeLlama \cite{codellama2} and DeepSeekCoder \cite{deepseek}. The models have been frequently investigated in prior work~\cite{deepseek,majdoub2024debugging,wang2024systematic}.  

Our selection includes models with distinct architectural or training features, making them well-suited for  code summarization. For example, the models are available in both instruction-tuned and non-instruction-tuned variants. Instruction-tuned models are optimized to process human-like instructions, making them particularly effective at manipulating natural language and code. This additional capability can be harnessed even in the context of QLoRA training, as demonstrated in prior work~\cite{yuan2023evaluating,fan2024exploring}.



\textit{CodeLlama}~\cite{codellama} is a family of open-source LLMs tailored for coding tasks. It is based on the general-purpose Llama-2 model \cite{llama2}, with further training on a corpus of 500B tokens that include both natural language and code. CodeLlama is available in several variants~\cite{codellamaHF}, each designed for specific use cases: a general-purpose coding model, an Instruct variant optimized for instruction tuning, and a Python-specialized version. The model sizes range from 7B to 70B parameters, and all versions are publicly accessible. CodeLlama has demonstrated strong performance in automating a range of code-related tasks \cite{zan2024codes,xia2023universal}, making it a representative model for our study. We used the general-purpose \emph{Instruct} version featuring 7B and 34B parameters in our experiments.



\textit{DeepSeek-Coder}~\cite{deepseek} is a set of open-source LLMs ranging from 1B to 33B parameters. These models are offered in two configurations: Instruct, optimized for instruction tuning, and Base. Trained on a dataset of two trillion tokens, including code-specific data, DeepSeek-Coder has been shown capable of outperforming larger models such as GPT-3.5  \cite{brown2020language}, while the small-sized version featuring 6.7B parameters has proven highly competitive to CodeLlama's 33B variant. For this study, we used the \emph{Instruct} version of DeepSeek-Coder in three variants, 1.3B, 6.7B, and 33B parameters. This selection served two different goals: 
 (i) it allowed us to compare the performance of QLoRA-optimized models against fully fine-tuned models by contrasting the results achieved by DeepSeek-Coder 1.3B in both configurations; and (ii) it enabled a comparison between small-sized (CodeLlama 7B \emph{vs.} DeepSeek-Coder 6.7B) and mid-sized models (\eg CodeLlama 34B \emph{vs}. DeepSeek-Coder 33B), providing insights into how QLoRA fine-tuning impacts performance across different model sizes.

\subsection{The QLoRA Fine-tuning Technique}
\label{sub:design_qlora}

QLoRA employs two innovative techniques for effective 4-bit finetuning: 4-bit NF4 quantization and Double Quantization, along with Paged Optimizers to manage memory efficiently during gradient checkpointing.


\subsubsection{NF4 Quantization}

The core of QLoRA’s approach lies in a method designed to efficiently quantize neural network weights into a 4-bit format which, uses NF4, a novel data type designed for AI applications. The 4-bit Normal Float~(NF4) data type is based on Quantile Quantization \cite{dettmers20218}, which ensures an even distribution of tensor values across quantization bins or categories. Using fast quantile approximation algorithms, QLoRA can estimate quantiles without the high computational costs associated with precise quantile calculations.
\looseness=-1

During this process, the neural network weights, which generally follow a zero-centered normal distribution, are adjusted to fit a predefined range. This normalization aligns the weight tensors with the range of the data type, allowing for more effective quantization by matching the tensor’s value distribution to that of the quantized format.


\subsubsection{Double Quantization}
To  further reduce memory footprint, QLoRA follows a two-step approach: (i) the model weights are quantized to 4-bit precision using NF4, and (ii) the quantization constants (scales and zero-points) from the first step are quantized to a lower precision. 
QLoRA implements Blockwise k-bit Quantization, where weights are divided into distinct blocks that are independently quantized, rather than quantizing all weights collectively. This method generates multiple quantization constants, which can undergo a second round of quantization, providing additional memory savings.
\looseness=-1

\subsubsection{Paged Optimizer}

When training large models, gradient checkpointing comes in handy as a technique to reduce memory usage during model training, yet memory spikes can still occur when processing mini-batch with a long sequence of input tokens.
Paged optimizers minimize GPU memory use by storing states in CPU memory and transferring them as needed.
\looseness=-1

\begin{figure}[t!]
	\centering
	\includegraphics[width=0.7\linewidth]{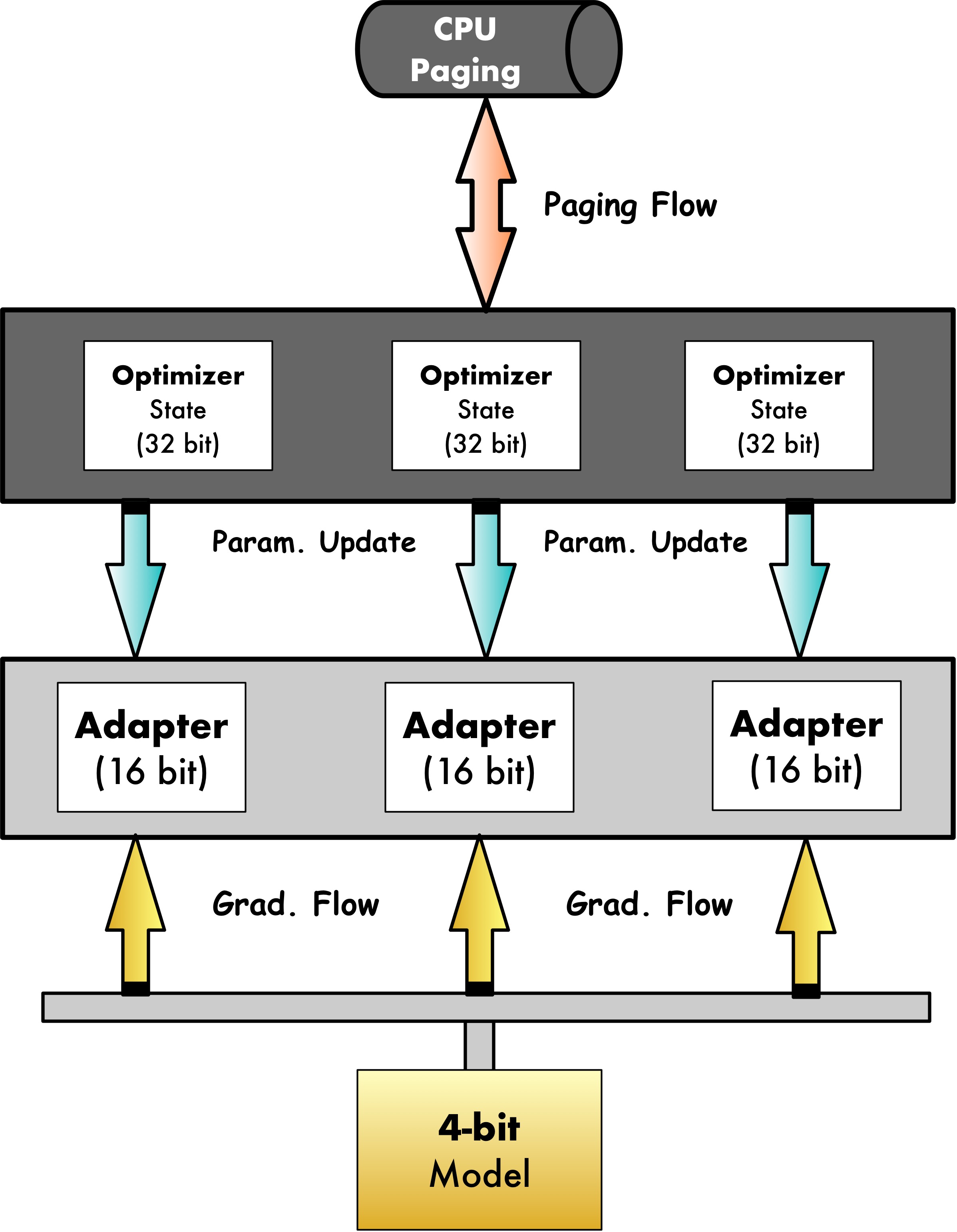}
	\caption{QLoRA finetuning with paged optimizers \cite{dettmers2024qlora}}
	\label{fig:qlora}
\end{figure}

\figref{fig:qlora} depicts the fine-tuning process of QLoRA, an extension of LoRA that, as noted, utilizes NF4 for efficient weight storage and BFloat16 for computations and gradient calculations. 
The addition of paged optimizer memory management further enhances efficiency, making QLoRA particularly suitable for resource-constrained environments.

In our study, we selected the QLoRA parameter configuration outlined in \tabref{tab:hp-tuning}, which includes three parameters: (i) \texttt{lora\_r}, (ii) \texttt{lora\_alpha}, and (iii) \texttt{lora\_dropout}. These were kept constant for each QLoRA fine-tuning instance.
The choice of the hyperparameter values follows established best practices to ensure precision while minimizing resource consumption \cite{dettmers2024qlora,hu2021lora}. 

\begin{table}[h!]
	\centering
	\caption{QLoRA hyperparameters used in our experiments}
	 \resizebox{.45\textwidth}{!}{
	\begin{tabular}{llcc}
		\hline
		& \bf  Parameter                   & \bf Description & \bf Value \\ 
		\hline
		& \texttt{lora\_r}       & \emph{lora attention dimension/ rank}   & 8 \\ 
		& \texttt{lora\_alpha}   & \emph{lora scaling parameter}           & 16 \\ 
		& \texttt{lora\_dropout} & \emph{lora dropout probability}        & 0.1 \\ 
		\hline
	\end{tabular}
	\label{tab:hp-tuning}
}
\end{table}

\smallskip

\subsection{Dataset and Model Training}
\label{sub:data}

We employed the \emph{Code-to-Text} dataset from the CodeXGLUE benchmark~\cite{codexglue, CodeXGLUEbench} to train and evaluate all QLoRA-optimized models, focusing specifically on Java and Python. The benchmark consists of pairs of code methods and their associated natural language descriptions, extracted from  $\sim$6 million instances of human-written code documentation.

Our decision to leverage CodeXGLUE was driven by its extensive use in prior research for studying LLMs in code-related tasks~\cite{wu2022learning,mastropaolo2022using,mastropaolo2021empirical,shi2022evaluation,chen2024code}. 

\tabref{tab:datasets} presents a summary of the datasets employed
for training and evaluation. To this extent, we train each QLoRA-optimized model using a fixed set of hyperparameters, as detailed in \tabref{tab:hp-tuning}. Each model was trained for \textit{10 epochs} with a consistent \textit{batch size of 32} maintained throughout all experiments. To prevent overfitting, we implemented an early stopping, saving a new checkpoint after every 5,000 training steps, and monitoring the performance of the models using the METEOR score, which acts as a highly reliable proxy for differences exceeding 2 points in evaluating the quality of code summaries as perceived by humans~\cite{roy2021reassessing}.
In particular, the training process stops if no improvements in the METEOR score are observed after 15K steps, which equals to a window of 3. This approach allowed us to effectively monitor model performance and ensured that we retained the best-performing checkpoint of the models for both programming languages.

\vspace{-5pt}
\begin{table}[h!]
	\centering
	\caption{\#Number of Data Instances in Training, Validation, and Testing Splits}
	\label{tab:datasets}
		\resizebox{0.8\columnwidth}{!}{
	\begin{tabular}{lrrr}
		\toprule
		\textbf{Language} & \textbf{Training} & \textbf{Validation} & \textbf{Testing} \\
		\midrule
		Java & 164,923 & 5,183 & 10,955 \\
		Python & 251,820 & 13,914 & 14,918 \\
		\bottomrule
	\end{tabular}
}
\end{table}

During training, each model takes as input tokens the tokenized code from the \texttt{code\_tokens} field in the JSON file, corresponding to either Java\footnote{\url{https://zenodo.org/record/7857872/files/java.zip}} or Python\footnote{\url{https://zenodo.org/record/7857872/files/python.zip}}. The output is the sequence of natural language tokens provided in the \texttt{docstring\_tokens} field and joined together to form a string, specific to each programming language. 
We configured the maximum sequence length to 300 tokens during the training stage based on our analysis of the token distribution for code and natural language in the Code-to-Text dataset.


In our implementation of QLoRA, we followed the findings from \cite{dettmers2024qlora} and applied QLoRA to all linear layers of the networks (\eg Feed-Forward Layers, Self-Attention Layers, and Projection Layers). This approach enables QLoRA to effectively adapt to the task at hand by leveraging the parameter space considered essential for optimal performance.

The training process for full fine-tuning followed the same configuration as used for QLoRA-optimized models, including batch size, number of training epochs, and early stopping.

\subsection{Metrics and Experimental Procedure}
\label{sub:analysis}


We started by fine-tuning DeepSeek-Coder 1.3B using QLoRA with the dataset described in \secref{sub:data}. As outlined in \secref{sub:design_llm}, we limited full fine-tuning (FFT) to the smaller model variants included in our study to mitigate the substantial computational costs associated with adjusting all parameters of larger models (\eg DeepSeek-Coder 33B).

We proceeded by evaluating the performance of the model when generating code summaries for Java and Python methods. To this end, we relied on metrics that have been widely used in prior code summarization research~\cite{zhang2022survey,zhang2020retrieval,leclair2020improved}: 
\looseness=-1
\begin{itemize}[itemindent=0.3cm,leftmargin=0cm,label=]
	\setlength\itemsep{0.1cm}
	
\item \textbf{BLEU} (BilinguaL Evaluation Understudy) \cite{bleu} measures the similarity between candidate (predicted) summaries and reference (oracle) summaries. This metric assesses the overlap of $n$-grams within the two summaries, ranging from 0 (completely dissimilar summaries) to 1 (identical summaries). We compute the BLEU score at the sentence level, fixing $n=4$.

\item \textbf{METEOR} (Metric for Evaluation of Translation with Explicit ORdering)~\cite{meteor} is computed as the harmonic mean of unigram precision and recall, with recall given a higher weight than precision. Unlike BLEU, METEOR utilizes stemming and synonym matching to align more closely with human judgments of sentence similarity. METEOR ranges from 0 to 1, with a value of 1 indicating two identical sentences.

\item \textbf{ROUGE} (Recall-Oriented Understudy for Gisting Evaluation)~\cite{lin2004rouge}  consists of a set of metrics for evaluating both automatic text summarization and machine translation methods. ROUGE metrics compare automatically generated summaries or translations against a set of reference summaries, typically authored by humans. In line with Roy \etal \cite{roy2021reassessing}, we computed ROUGE-N(1-4), ROUGE-L, and ROUGE-W. ROUGE-N measures the number of matching $n$-grams between the generated summary and the reference summary with results reported in terms of recall, precision, and F1-score.

\item \textbf{chrF} (character $n$-gram F-score)~\cite{popovic2015chrf} measures the similarity between generated and reference summaries at the character level (rather than at the token level, done by the above metrics), reporting  a F1-score value.

\item \textbf{BERTScore}~\cite{zhang2019bertscore} computes sentence similarity using the embedding of a BERT model \cite{devlin2018bert} trained on English textual data. We report the F1-score (BERTScore-F1).

\item \textbf{SIDE} (Summary Alignment to Code Semantics)~\cite{mastropaolo2024evaluating} offers an automated method for evaluating the alignment between a Java method and its corresponding summary. It produces a score within the range of [-1, 1], where values closer to 1 indicate a stronger alignment between the comment and the documented code component. Conversely, lower SIDE scores signify weaker alignment, highlighting discrepancies between the summary and the code.
\end{itemize}

Next, we conducted a comprehensive end-to-end, task-specific fine-tuning of DeepSeek-Coder, updating 1.3 billion parameters. This process, backpropagates gradients through the entire model, enabling optimal adjustment of each parameter to enhance task-specific performance (\ie code summarization).
\looseness=-1

Finally, we assessed whether there are statistically significant differences in performance between fully fine-tuned models and QLoRA-optimized models. To this extent, we employed the Wilcoxon signed-rank test \cite{wilcoxon}, and measured the effect size using Cliff’s Delta (d) \cite{Cliff:2005}. The effect sizes are categorized as follows: negligible if $|d| < 0.10$, small if 0.10 $\leq$ $|d| < 0.33$, medium if 0.33 $\leq$ $|d| < 0.474$, and large if $|d|$ $\geq$ 0.474. We used a 95\% significance level across all tests and, since we tested our hypotheses through multiple tests, we adjusted the \emph{p}-values using Holm’s correction procedure~\cite{Holm1979a}.
The tests were computed for every metric included in our evaluation.
\looseness=-1


To investigate the impact of various sizes of models, we fine-tuned four model variants with QLoRA optimization: CodeLlama 7B/34B and DeepSeek-Coder 6.7B/33B. Next, we evaluated the performance of each model configuration in generating meaningful code descriptions for Python and Java methods. Additionally, we applied the Wilcoxon signed-rank test to analyze performance differences in evaluation metrics across models of varying sizes.

Each experiment was performed on a server running Ubuntu 22.04.5 LTS (GNU/Linux 5.15.0-125-generic x86\_64), equipped with two Nvidia L40S GPUs, each featuring 48GB of graphics memory.

\subsection{Qualitative Analysis}
\label{sec:qualitative}

\begin{figure}[t]
	\centering
	\includegraphics[width=\columnwidth]{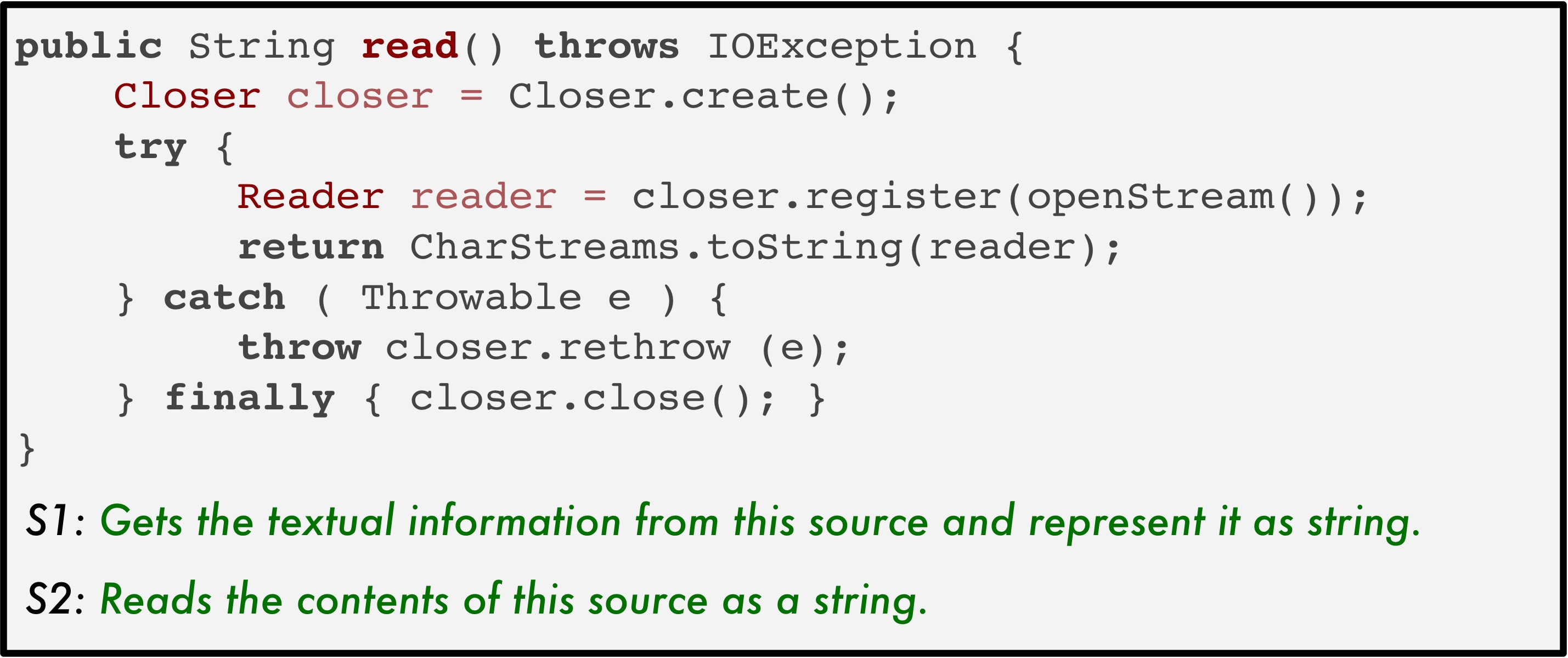}
	\caption{Semantically equivalent Java code summaries.} 
	\label{fig:semantically-equivalent}
\end{figure}

Evaluating code summarization can present unique challenges, particularly in dealing with semantically equivalent summaries. For example, in \figref{fig:semantically-equivalent}, the two summaries indicated with $S1$ and $S2$, describe the same functionality of a code snippet but use different phrasing. This discrepancy poses a challenge because traditional evaluation metrics like BLEU or ROUGE rely heavily on exact word matches and may penalize the generated summary $S1$ for not being identical to the ground truth $S2$. In addition, Mastropaolo \etal \cite{mastropaolo2024evaluating} have recently demonstrated that word-overlap metrics like BLEU, and even embedding-based metrics such as BERTScore \cite{zhang2019bertscore}, can only capture one of the several dimensions pertaining to the evaluation of code summarizers.
Thus, to provide a more accurate assessment of the capabilities of QLoRA in fine-tuning models for bi-modal software engineering tasks, we manually analyzed two statistically significant, randomly selected samples: one consisting of 384 Java methods and the other of 384 Python methods, generated by the best-performing QLoRA-optimized model identified in our study.

With this manual analysis, we aimed to better understand the nuances in code summarization tasks that automated metrics might miss, offering a comprehensive evaluation of how effectively QLoRA-optimized models can support bi-modal SE-related tasks across two programming languages.

For this analysis, two paper authors  independently reviewed the 768 incorrectly generated summaries, equally split between Python and Java (\ie 384 + 384). Any conflicts were resolved through open discussion between the reviewers, involving a third author when needed. The summaries were sampled from the set of incorrect predictions made by CodeLlama-34B, the best-performing model. Each prediction was classified as:

\begin{itemize} 
	\item  \emph{Semantically equivalent}: The ground truth and the model’s prediction use different wording but convey exactly the same information to the developer.
	
	\item \emph{Meaningful code description}: These cases represent instances where the model generated a code summary that not only conveyed the intended information but was of better quality than the ground truth.
	A few examples are reported in \figref{fig:qualitative-examples}.
	
	\item \emph{Partially equivalent}: The prediction includes only part of the information conveyed in the ground truth. While these predictions can still be useful for the developer, some code adjustments are needed to align the prediction with the original method.
	
	\item \emph{Incorrect}: The code summary predicted by the model is documenting something else and not the underlying code.

\end{itemize}

To assess inter-rater reliability among the evaluators, we calculated Krippendorff’s $\alpha$ coefficient \cite{krippendorff2011computing}. For the analysis of Java elements, the $\alpha$ coefficient was 0.752, and for Python, it was 0.803. In both cases, the $\alpha$ that ranges between [-1;1] indicated a high level of agreement between the two evaluators. Following this, we report the percentage of instances in each category detailed above.

\section{Results and Discussion}
\label{sec:results}


In this section, we present and discuss the results of our study addressing our RQ, which aims to evaluate the effectiveness and memory footprint of the CLMs when fine-tuned with QLoRA, compared to full fine-tuning.  

\subsection{QLoRA vs Full Fine-Tuning}

As described in \secref{sub:analysis}, we fine-tuned DeepSeek-Coder 1.3B under two configurations: QLoRA fine-tuning and full fine-tuning (FFT). The CodeLlama models are only fine-tuned with QLoRA due to the excessive computational costs associated with full fine-tuning, which we could not afford. \tabref{tab:parameters} illustrates the numbers of \textit{total model parameters} and \textit{trainable parameters} for our selected models. 

\begin{table*}[t]
	\centering
	\caption{Performance of different QLoRA Fine-tuned parameter size models. The best results across all the experimented configurations are highlighted in yellow.}
	\scriptsize
	\label{tab:model-per}
	\resizebox{\linewidth}{!}{
		\begin{tabular}{lllccccccc}
			\toprule
			\textbf{Model} & \textbf{Parameter Size} & \textbf{Training} & \textbf{Dataset Type} & \textbf{BLEU} & \textbf{METEOR} & \textbf{Rouge-L} & \textbf{chrF} & \textbf{BERTScore F1} & \textbf{SIDE} \\
			\midrule
			 \multirow{4}{*}{\textbf{\textit{CodeLlama}}}

			& \multirow{2}{*}{7B} & \multirow{2}{*}{QLoRA} & 
			\cellcolor[gray]{.70}   Python & \cellcolor[gray]{.70} 8.9\% & \cellcolor[gray]{.70}  33.9\% &  \cellcolor[HTML]{FFDEAD}  \cellcolor[gray]{.70}  \cellcolor[HTML]{FFDEAD} \bf 36.5\% & \cellcolor[gray]{.70}  29.7\% &  \cellcolor[gray]{.70} 88.8\% & \cellcolor[gray]{.70} -  \\
			& & & \cellcolor[gray]{.9} Java & \cellcolor[gray]{.9}10.3\% & \cellcolor[gray]{.9} 35.8\% & \cellcolor[gray]{.9} \cellcolor[HTML]{FFDEAD} \bf 37.4\% &\cellcolor[gray]{.9}  30.9\% & \cellcolor[gray]{.9}  89.0\% & \cellcolor[gray]{.9} 87.5\%  \\
			
			\cline{2-10}

			& \multirow{2}{*}{34B} & \multirow{2}{*}{QLoRA} & \cellcolor[gray]{.70} Python & \cellcolor[gray]{.70}   \cellcolor[HTML]{FFDEAD} \bf 10.7\% & \cellcolor[gray]{.70} \cellcolor[HTML]{FFDEAD} \bf 35.3\% & \cellcolor[gray]{.70} 35.2\% & \cellcolor[gray]{.70}   \cellcolor[HTML]{FFDEAD} \bf 31.3\% & \cellcolor[gray]{.70}   89.0\% & \cellcolor[gray]{.70} - \\

			& & & \cellcolor[gray]{.9} Java & \cellcolor[gray]{.9} \cellcolor[HTML]{FFDEAD} \bf 11.6\% &\cellcolor[gray]{.9} \cellcolor[HTML]{FFDEAD} \bf 37.2\% & \cellcolor[gray]{.9} 36.7\% & \cellcolor[gray]{.9}  \cellcolor[HTML]{FFDEAD} \bf 32.0\%  & \cellcolor[gray]{.9}   \cellcolor[HTML]{FFDEAD} \bf 89.1\% & \cellcolor[gray]{.9} 86.6\%  \\
			
			\midrule
			
			\multirow{8}{*}{\textbf{\textit{DeepSeek-Coder}}}        
			& \multirow{4}{*}{1.3B} & \multirow{2}{*}{\bf FFT} & \cellcolor[gray]{.70} Python & \cellcolor[gray]{.70} 6.5\% & \cellcolor[gray]{.70} 32.3\% & \cellcolor[gray]{.70} 30.1\% & \cellcolor[gray]{.70} 27.4\% & \cellcolor[gray]{.70} 88.2\% & \cellcolor[gray]{.70} - \\
			& & & \cellcolor[gray]{.9} Java & \cellcolor[gray]{.9} 8.2\% & \cellcolor[gray]{.9} 33.4\% & \cellcolor[gray]{.9} 32.2\% & \cellcolor[gray]{.9} 28.8\% & \cellcolor[gray]{.9} 88.4\% & \cellcolor[gray]{.9} 87.3\% \\
			\cline{3-10}
			
			& & \multirow{2}{*}{QLoRA} & \cellcolor[gray]{.70} Python & \cellcolor[gray]{.70} 7.4\% & \cellcolor[gray]{.70} 34.2\% & \cellcolor[gray]{.70} 32.0\% & \cellcolor[gray]{.70} 28.0\% & \cellcolor[gray]{.70} 88.5\% & \cellcolor[gray]{.70} - \\
			& & & \cellcolor[gray]{.9} Java & \cellcolor[gray]{.9} 8.7\% & \cellcolor[gray]{.9} 35.1\% & \cellcolor[gray]{.9} 33.9\% & \cellcolor[gray]{.9} 29.0\% & \cellcolor[gray]{.9} 88.6\% & \cellcolor[gray]{.9}  \cellcolor[HTML]{FFDEAD} \bf 88.1\%  \\
			\cline{2-10}
			
			& \multirow{2}{*}{6.7B} & \multirow{2}{*}{QLoRA} & \cellcolor[gray]{.70} Python & \cellcolor[gray]{.70} 8.6\% & \cellcolor[gray]{.70} 36.5\% &  \cellcolor[gray]{.70} \cellcolor[gray]{.70}\cellcolor[gray]{.70} 33.9\% & \cellcolor[gray]{.70} 29.4\% & \cellcolor[gray]{.70} 88.8\% & \cellcolor[gray]{.70} -  \\		
			& & & \cellcolor[gray]{.9} Java & \cellcolor[gray]{.9} 9.9\% & \cellcolor[gray]{.9} 37.3\% & \cellcolor[gray]{.9} 35.5\% & \cellcolor[gray]{.9} 30.1\% & \cellcolor[gray]{.9} 89.0\% &  \cellcolor[gray]{.9} 87.9\% \\
			\cline{2-10}

			& \multirow{2}{*}{33B} & \multirow{2}{*}{QLoRA} & \cellcolor[gray]{.70} Python & \cellcolor[gray]{.70} 10.5\% & \cellcolor[gray]{.70} 37.8\% & \cellcolor[gray]{.70} 35.2\% & \cellcolor[gray]{.70} 31.2\% & \cellcolor[gray]{.70}  89.0\% &  \cellcolor[gray]{.70} - \\
			& & & \cellcolor[gray]{.9} Java & \cellcolor[gray]{.9} 10.9\% & \cellcolor[gray]{.9} 38.1\% & \cellcolor[gray]{.9} 36.4\% & \cellcolor[gray]{.9} 31.5\% & \cellcolor[gray]{.9}  89.0\% & \cellcolor[gray]{.9} 87.4\%  \\
			
			\\

			\bottomrule

		\end{tabular}
	}
	\vspace{-0.2cm}
\end{table*}

\begin{table*}[t]
	\centering
	\caption{Summary of Model Training Parameters and Memory Utilization in Megabytes (MB)}
	\scriptsize
	\label{tab:parameters}
	\resizebox{\linewidth}{!}{
		\begin{tabular}{llccccccc}
			\toprule
			\bf Training Strategy & \bf Model & \bf Peak GPU Mem. Consumption (MB) & \bf Trainable Parameters & \bf Model's Parameters & \bf Trainable \% \\
			\midrule
			\multirow{5}{*}{\emph{QLoRA Fine-Tuning}} & CodeLlama-7b & 11,877  & 20,277,376 & 6,758,820,064 & 0.300 \\
			& CodeLlama-34b & 37,424 & 54,781,952 & 33,798,752,256 & 0.162 \\
			
			\noalign{\smallskip}
			\cline{2-6}
			\noalign{\smallskip}
			& DeepSeek-Coder-1.3b &    5,154  & 7,770,112 & 1,354,242,048 & 0.574 \\
			& DeepSeek-Coder-6.7b & 12,894  & 20,279,296 & 6,760,792,064 & 0.300 \\
			& DeepSeek-Coder-33b &  39,724 & 61,898,752 & 33,404,890,112 & 0.185 \\
			\midrule
			\emph{Full Fine-Tuning} & DeepSeek-Coder-1.3b & 16,776  & 1,354,242,048 & 1,354,242,048 & 100 \\
			\bottomrule
		\end{tabular}
	}
	\vspace{-0.2cm}
\end{table*}



\tabref{tab:model-per} summarizes the performance results for the five models optimized using QLoRA, alongside the Full Fine-Tuning (FFT) of DeepSeek-Coder 1.3B. A key insight that emerges after observing the second and third rows is that QLoRA-based fine-tuning consistently delivers superior performance compared to full fine-tuning of DeepSeek-Coder 1.3B across the two programming languages.
For example, in terms of the METEOR score, the QLoRA fine-tuned model surpasses its fully fine-tuned counterpart by approximately 2\% for both Python and Java. Similarly, based on the ROUGE-L metric, performance improvements range from 1.9\% to 2.7\%, with the largest gains observed for Java. For each metric reported in \tabref{tab:model-per}, the observed differences were found to be statistically significant (based on a Wilcoxon signed-rank test at 95\% significance), though the effect size is negligible. The detailed results of this analysis are available in our online replication package~\cite{replication}.

\tabref{tab:parameters} shows the results of our analysis of the peak GPU memory consumption during model fine-tuning. The table reveals a substantial difference between QLoRA and full fine-tuning. FFT of the 1.3B parameters requires an average of approximately 16GB of GPU memory, while QLoRA fine-tuning significantly reduces memory usage, requiring only 5GB on average--saving 10GB compared to FFT (\ie about a third of the FFT memory footprint). 
This reduction in memory consumption can be attributed to the considerably lower number of trainable parameters in QLoRA (\tabref{tab:parameters}--Columns 4 and 5), which represents a significant scale difference (millions vs. billions) compared to training all parameters. The freed-up memory can be used to support additional tasks, enable larger batch sizes, or improve overall computational efficiency, making QLoRA a more practical and scalable solution for resource-constrained environments.

The results indicate that QLoRA is not only effective in optimizing resource efficiency but also in outperforming full fine-tuning. These findings align with those presented in the pioneering work that introduced QLoRA~\cite{dettmers2024qlora}. 
 
In the software engineering literature, Weyssow \etal~\cite{weyssow2023exploring} demonstrated that for coding activities--particularly code generation--the use of LoRA adapters for fine-tuning large language models outperforms FFT. 
Furthermore, in the same study, they showed that QLoRA fine-tuning surpasses LoRA fine-tuning for the same task, establishing a clear hierarchy of fine-tuning strategies: FFT $<$ LoRA fine-tuning $<$ QLoRA fine-tuning. Hence, if LoRA fine-tuning already outperforms FFT, this suggests that once the underlying model’s knowledge, distilled within its parameters, is frozen--as in LoRA and QLoRA fine-tuning--adapting a smaller subset of parameters is sufficient to effectively capture the nuances of the intended task.
As we show, QLoRA improves model performance while reducing memory footprint.
This result might seem counterintuitive, given that QLoRA relies on extreme quantization to optimize memory usage. However, a possible explanation for this behavior comes from Dettmers \etal (the authors of QLoRA), who observe that any performance degradation due to information loss during quantization is not only fully recovered but often surpassed through the fine-tuning of LoRA modules after the quantization process. 

\vspace{-3pt}
\begin{boxM}
	\textbf{\emph{Finding$_{1}$}:} QLoRA fine-tuning for code summarization delivers performance on par with what is observed for other software engineering tasks such as code generation \cite{weyssow2023exploring}.
	By optimizing a limited subset of quantized parameters, it outperforms full fine-tuning in terms of predictive performance and memory consumption.
\end{boxM}

Examining the impact of varying parameter counts, the results align with our expectations: larger models consistently outperform their smaller counterparts when QLoRA is applied to support code summarization tasks (see \tabref{tab:model-per}). For example, CodeLlama-34B demonstrates significantly higher performance compared to its 7-billion-parameter variant, and a similar pattern is found with DeepSeek-Coder, where larger versions achieve superior results. These trends hold true across both programming languages, underscoring the effect of QLoRA regardless of the model size.
However, this improvement comes at a cost. Larger models demand significantly more GPU memory during training, with usage peaking at 40 GB when fine-tuning DeepSeek-Coder 33B and 37.5 GB when fine-tuning CodeLlama 34B, as detailed in \tabref{tab:parameters}.
This highlights the trade-off between model size, performance, and resource requirements. Depending on the final application and available hardware, one may need to prioritize performance over resource consumption or vice versa.
In other words, if GPU memory allocation is limited, sacrificing some performance may be a reasonable trade-off, especially considering the capabilities of models with around 7B parameters.
The observed improvements resulted in statistically significant differences, though the effect sizes of these differences are negligible. 
The detailed results of all the statistical analyses are available in our online replication package~\cite{replication}.
Scaling up to DeepSeek-Coder 33B, the improvements, while still statistically significant, exhibit diminishing returns. The performance gap narrows across all evaluated metrics, as evidenced by the negligible effect sizes. Although larger models generally offer greater capacity, the diminishing improvements suggest that further scaling might not always justify the increased computational resources, particularly for code summarization.

In contrast, for CodeLlama, scaling from 7B to 34B yields more pronounced gains. The larger 34B variant achieves a 3-4\% improvement in METEOR scores over its smaller counterpart, equating to an 11.8\% improvement for Python and  8.4\% for Java. These results highlight the effectiveness of both the CodeLlama model family and DeepSeek-Coder in leveraging increased parameter counts to enhance performance. This trend becomes particularly apparent when examining the top-performing models in \tabref{tab:model-per}, where the best results across all experimental configurations are highlighted in yellow. Notably, four out of six metrics in our evaluation reach their highest values with models from the CodeLlama family that have a parameter count exceeding 30B.
The performance differences remain consistent across embedding-based metrics (BERTScore-F1 for both languages, SIDE for Java) and are statistically significant but with negligible effect sizes, suggesting limited practical impact. This reinforces that while larger models can improve performance, the gains may not justify the increased GPU memory consumption.

\begin{boxM}
	\textbf{\emph{Finding$_{2}$}:}  Larger models generally offer greater capacity and potential better support for code summarization, but they eventually reach a point of diminishing return. However, if maximizing performance is the primaryobjective, CodeLlama 34B delivers the best outcomes
	for code summarization, with improvements of 11.8\% for Python and 8.4\% for Java.
\end{boxM}



\subsection{Qualitative Analysis}

\begin{figure}[t!]
	\centering
	\includegraphics[width=\columnwidth]{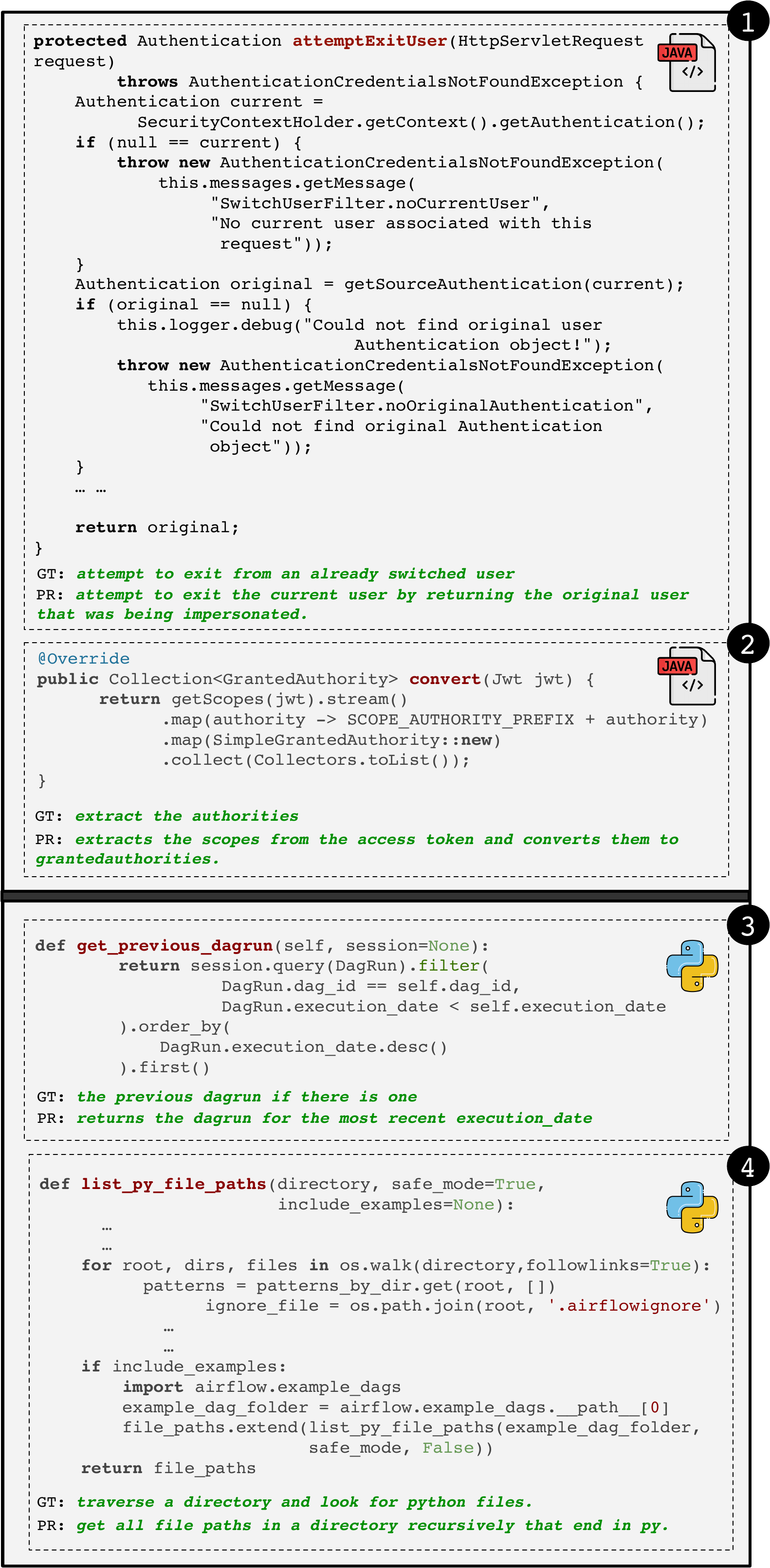}
	\caption{Examples of predictions made by CodeLlama 34B that have been labeled as \emph{meaningful code summaries}. }
	\label{fig:qualitative-examples}
\end{figure}

\figref{fig:qualitative-examples} presents four triplets of $\langle$Method, Target$_{summary}$ or GT, Predicted$_{summary}$ or PR$\rangle$. The examples are divided by programming language, with two for each language. The top section features Java examples, while the bottom showcases two Python examples. In all cases, the model-generated comments are not only accurate compared to the ground truth but also offer improvements over it. We remind the reader that this type of code comment falls under the category of \emph{meaningful code descriptions} (\secref{sec:qualitative}).

Specifically, focusing on the first triplet~\circled{1}, the developer-provided ground truth summary, \texttt{attempt to exit from an already switched user}, encapsulates the method's basic functionality. However, the CodeLlama 34B model optimized  using QLoRA, generates a summary that clarifies the method’s logic, explicitly documenting that the method \texttt{attempts to exit the current user by returning the original user that was being impersonated}. This provides valuable additional information, specifically noting that \texttt{the method returns the original user who was being impersonated}. This distinction makes the summary more comprehensive, as it clarifies the intended logic and provides insights into the rationale of the code.


In \circled{2}, the predicted summary demonstrates significant improvement, as the model captures details that the developer overlooked or deemed unnecessary, such as the process of extracting tokens. The model not only identifies these elements but also elaborates on them, providing a more comprehensive and actionable summary: \texttt{extracts the scope from the access token and converts them to grant authorities}. This enhanced prediction demonstrates the model’s capability to infer additional context and generate summaries that not only ``copy'' tokens already present in the method but also synthesize new information. For instance, the word-token \texttt{access} was inferred through the model’s deeper understanding of the code’s logic and intent. This ability allows for the creation of more comprehensive and insightful summaries.

Turning to the Python first example \circled{3}, the prediction adds depth by conveying an additional message: that the \texttt{dagrun must be retrieved based on the most recent execution date}.  This enriched context provides developers with summaries that are not only concise but also contextually informative.
The second Python example (\circled{4}) illustrates that, despite the significantly smaller number of parameters adjusted during fine-tuning (\ie millions rather than billions), the performance of the QLoRA-optimized model remains, even for tasks demanding high contextual reasoning. Notably, the model identifies an important detail revealed only at the conclusion of the method: \emph{list\_py\_paths is a recursive method}. This ability to detect nuanced information demonstrates the model’s effectiveness in generating meaningful and context-aware summaries despite the limited parameter adjustment, quantized and dequantized, as explained in \secref{sub:design_qlora}.

These instances align with the findings of the manual investigation conducted on 384 incorrect Java summaries and 384 incorrect Python summaries.

For Java, 15.36\% of conflicts arose during the labeling process. These conflicts, resolved by a third author who was not involved in the initial labeling, resulted in the following breakdown: 31.07\% of summaries were deemed semantically equivalent to the ground truth, 53.0\% were partially equivalent, and in 8.87\% of cases, CodeLlama 34B provided summaries that were more accurate and informative than those written by developers. Finally, 7.57\% of the summaries were classified as incorrect.
\looseness=-1

A similar pattern was observed for Python, with slight variations in the distribution across categories. Specifically, CodeLlama 34B generated 37.86\% of summaries as a real developer would do (\ie semantically equivalent summaries), 53.0\% were partially equivalent, and in 3.66\% of cases, CodeLlama 34B produced recommendations superior to those of developers. Finally, 5.48\% of the recommendations were found to be incorrect.

\subsection{Can We Translate the QLoRA Benefits for Code Summarization to General-Purpose Language Models?}
\label{sub:phi3}

Our investigation demonstrated that QLoRA can serve as a resource-efficient training strategy for code summarization, paving the way for advancements in Code-to-NL tasks. However, an open question remains: \emph{can these findings be generalized to models that have been pre-trained—not solely but partially—on software engineering data?}

The rationale behind this question lies in the growing adoption of hybrid models, such as Phi-3 mini \cite{abdin2024phi}, which have been benchmarked extensively on coding tasks while also serving as baselines in comparisons against both code-specific and general-purpose language models \cite{deng2024assessing}. These models blur the lines between domain-specific and general-purpose architectures, offering a unique ground for evaluating the transferability of QLoRA's fine-tuning benefits.

For such analysis, we selected Phi-3 mini \cite{abdin2024phi}, a 3.8-billion-parameter model introduced by Microsoft in 2024 that was shown to achieve results on par with Llama 3 \cite{dubey2024llama}, Meta’s state-of-the-art 7-billion-parameter model, when applied to code-related tasks such as code generation. Phi-3 mini is nearly half the size of Llama 3 \cite{dubey2024llama}. We chose this Microsoft model for the generalizability analysis due to its popularity  and its frequent use as a baseline for coding tasks \cite{deng2024assessing, peixoto2024effectiveness}.

\begin{table*}[h!]
	\centering
	\caption{Performance of phi-3-mini when Fully Fine-tuned and QLoRA Fine-tuned}
	\scriptsize
	\label{tab:phi3}
	\resizebox{0.90\linewidth}{!}{
		\begin{tabular}{lllccccccc}
			\toprule
			
			\textbf{Model} & \textbf{Parameter Size} & \textbf{Training} & \textbf{Dataset Type} & \textbf{BLEU} & \textbf{METEOR} & \textbf{Rouge-L} & \textbf{chrF} & \textbf{BERTScore F1} & \textbf{SIDE} \\
			\midrule

			\multirow{4}{*}{\textbf{\textit{Phi-3-mini}}}        
			& \multirow{4}{*}{3.8B} &  \multirow{2}{*}{ \bf FFT} & \cellcolor[gray]{.70} Python & \cellcolor[gray]{.70} 6.1\%  & \cellcolor[gray]{.70} 31.8\%  & \cellcolor[gray]{.70} 29.4\%  &  \cellcolor[gray]{.70} 26.3\% & \cellcolor[gray]{.70}  88.1\%  & \cellcolor[gray]{.70} -   \\
			& & & \cellcolor[gray]{.9} Java & \cellcolor[gray]{.9} 7.9\%  & \cellcolor[gray]{.9} 33.6\%  & \cellcolor[gray]{.9} 32.1\%  & \cellcolor[gray]{.9} 27.7\% & \cellcolor[gray]{.9} 88.5\%  &  \cellcolor[gray]{.9} 87.3\% \\
			
			\cline{3-10}
			
			& & \multirow{2}{*}{QLoRA} & \cellcolor[gray]{.70} Python & \cellcolor[gray]{.70}  7.9\% & \cellcolor[gray]{.70} 35.4\%  & \cellcolor[gray]{.70} 32.9\%  & \cellcolor[gray]{.70} 28.5\% & \cellcolor[gray]{.7} 88.7\%  &  \cellcolor[gray]{.70} -  \\
			& & & \cellcolor[gray]{.9} Java & \cellcolor[gray]{.9} 8.7\% & \cellcolor[gray]{.9} 35.9\%  & \cellcolor[gray]{.9} 34.4\%  & \cellcolor[gray]{.9} 28.9\% & \cellcolor[gray]{.9} 88.8\% &  \cellcolor[gray]{.9} 88.1\% 
			\\
		
			\bottomrule

		\end{tabular}
	}
	\vspace{-0.2cm}
\end{table*}

We trained Phi-3 mini using two distinct configurations, replicating the approach used for DeepSeek-Coder 1.3B (\secref{sec:design}). Specifically, we conducted both full model fine-tuning and QLoRA fine-tuning, followed by model evaluation in each scenario. The training and evaluation processes were carried out as outlined in \secref{sec:design}.

\tabref{tab:phi3} presents the results of our experiments with Phi-3 mini. Notably, focusing on the BLEU metrics, the QLoRA-optimized model consistently outperforms the fully fine-tuned model, showing a performance improvement of 2\% for Python and 0.9\% for Java. This trend extends to other metrics, such as METEOR, ROUGE-L, and chrF, where the QLoRA-optimized model surpasses the FFT model by a margin of approximately 2–3\%.
Further analysis compares the performance of code models with the general-purpose phi-3 mini model. While code models demonstrate better performance than Phi-3 mini, the observed gap is not substantial. To validate this finding, we conducted a Wilcoxon signed-rank test between the Phi-3 mini-3.8B and DeepSeek-Coder-1.3B models, evaluating both fully fine-tuned and QLoRA fine-tuned versions.
The analysis failed to reveal statistically significant differences  for Java (regardless of the evaluation metric), whereas, for Python, this behavior was observed only for the BLEU metric in the fully fine-tuned setup. These analysis details are included in our replication package~\cite{replication}.
\vspace{5pt}

\section{Implications of our findings} 
\label{sec:implications}


Among the various applications of code-related bi-modal tasks, code summarization stands as a fundamental endeavor in software engineering, playing a crucial role in enhancing developer productivity \cite{majdoub2024debugging, mastropaolo2021studying}, improving code comprehension~\cite{zhang2020retrieval, fang2024esale}, and supporting software maintenance \cite{yang2024multi}.

Based on our findings, we draw the following implications:

\smallskip

\noindent\textbf{Improving CLMs Sustainability via QLoRA fine-tuning.} By applying QLoRA fine-tuning to two SoTA code models for code summarization activities, we demonstrated that it achieves competitive results compared to full model fine-tuning. Additionally, QLoRA significantly reduces memory requirements, cutting the memory footprint by approximately a third (see \tabref{tab:parameters}), which enhances the scalability, sustainability, and usability of advanced AI systems built on large language models for code. 


\smallskip

\noindent\textbf{QLoRA streamlines CLMs training for code summarization with comparable success to code generation.}\\
The successful adaptation of QLoRA for code summarization underscores its potential to enhance a wide range of code-related tasks through efficient fine-tuning.
In software engineering research, this paves the way for exploring QLoRA's capability to fine-tune CLMs for hybrid tasks that integrate code and natural language, such as code review.
Examining how QLoRA performs in these hybrid scenarios could unlock new possibilities for leveraging code models in complex, real-world applications. 

\section{Threats to Validity}
\label{sec:threats}

\textbf{Construct Validity} threats pertain to the relationship between theory and observation, primarily concerning the measurements used to answer our research questions. In this context, the main threat in our study lies in the selection of metrics to evaluate the quality of the generated summaries. As outlined in \secref{sub:analysis}, we utilized well-established metrics for assessing code summary quality \cite{haque2022semantic}, along with newer metrics such as SIDE \cite{mastropaolo2024evaluating}.

\textbf{Internal Validity} threats relate to factors internal to our study that could affect the achieved results. One possible threat can be the selected models to conduct the analysis. We partially mitigate this issue by considering models from the state-of-the-art that have been used in several prior studies~\cite{virk2024enhancing,zhu2024effectiveness,yang2024synthesizing,li2023structured}. 


Another threat is the selection of the QLoRA hyperparameters. In this regard, we resorted to the literature and particularly followed the recommendations provided in the original paper that introduced QLoRA \cite{dettmers2024qlora}. 

A further potential limitation concerns memory measurement, which was conducted based on a single run within a specific environment. Factors such as background processes or system load variations could impact memory usage metrics. To mitigate this, we maintained a controlled and consistent environment across experiments. However, we acknowledge that conducting multiple runs under varying conditions could offer additional insights.

\textbf{Conclusion Validity} threats involve the link between the experimental process and its outcomes. As detailed in \secref{sub:analysis}, where appropriate, our conclusions are supported by suitable statistical methods.

\textbf{External Validity} threats concern the generalizability of our findings.  In this regard, our study focuses solely on code summarization as a representative bi-modal task. Furthermore, although we conducted experiments using Python and Java— widely studied programming languages in software engineering automation \cite{hou2023large,deepseek,codellama}—we recognize that results may differ for other programming languages.

\section{Conclusions and Future Work} 
\label{sec:conclusions}

This study found that QLoRA is as effective for code summarization as it is for code generation---a task similar in intent but distinct in execution. Specifically, our findings show that QLoRA not only matches but consistently outperforms full model fine-tuning, delivering superior results while significantly improving resource efficiency. Notably, QLoRA excels in memory utilization, positioning it as a practical solution for resource-constrained environments.

Building on the positive impact of QLoRA fine-tuning for complex code-related tasks, our future research will focus on achieving a balance between efficiency and performance when deploying QLoRA in practical applications, such as live coding assistance tools.


\balance
\bibliographystyle{IEEEtran}
\bibliography{main}

\end{document}